# Two-Dimensional GaN: An Excellent Electrode Material Providing Fast Ion Diffusion and High Storage Capacity for Li-Ion and Na-Ion Batteries

Xiaoming Zhang, Lei Jin, Xuefang Dai, Guifeng Chen, and Guodong Liu*

School of Materials Science and Engineering, Hebei University of Technology, Tianjin 300130, PR China.
E-mail:gdliu1978@126.com

**ABSTRACT:** Identifying applicable anode materials is a significant task for Li- and Na-ion battery (LIB and NIB) technologies. We propose the GaN monolayer (2D GaN) can be a good anode candidate. The GaN monolayer manifests stable Li/Na adsorption and inherently low theoretical voltages. Most excitingly, both high storage capacity and extremely fast Li/Na diffusion can be simultaneously realized in the GaN monolayer. For Li, the storage capacity and diffusion barrier is 938 mA h $g^{-1}$ and 80 meV , respectively. And the values for Na are 625 mA h $g^{-1}$ and 22 meV. Comparing with known 2D anode materials with similar scale of ion diffusion barriers, the GaN monolayer almost possesses the highest Li/Na storage capacity discovered to date. Our work suggests that the 2D GaN is a prospective anode material offering fast ion diffusion and high storage capacity.

# 1 Introduction

The Li-ion batteries (LIBs) have been intensively used nowadays. The success of LIBs mainly benefits from their excellent energy conversion efficiency and good storage capacity.[1-5] Nevertheless, the development of LIBs has currently faced increasing challenge due to the naturally limited lithium resource and their high production cost.[6-8] As a result, non-Li-ion batteries have emerged in recent years.[9-15] Among them, Na-ion batteries (NIBs) have attracted the most attentions.[9,10] This arises from that the sodium has relatively low cost. For LIBs and NIBs, the electrochemical performance is the most dependent on their electrode materials; thus developing good electrode materials is one of major focuses in current battery technologies. Two-dimensional (2D) materials have offered great potential as ion battery electrodes, because their fully exposed surfaces are believed to provide fast ion diffusion and the maximum ion insertion channels.[16-18] Up to date, several families of 2D materials, such as graphene,[19-21] MXenes,[22-28] transition-metal dichalcogenides (TMDCs),[29-32] transition-metal dinitrides (TMDNs),[33] borophenes,[34-36] and others[37-40] have been identified as LIBs and NIBs electrodes. For battery electrode materials, the ion diffusion speed and the storage capacity are two of the most crucial indicators for the electrode performances. However, most known 2D electrode materials can only achieve high performance on one of these indicators. 2D anode materials, for example, $Mo_2C$,[24,25] $Nb_2C$[26] and $Ti_3C_2$[23] are proposed to offer very fast Li/Na diffusion speed but have low or moderate storage capacities; silicence[41] and $\beta_{12}/\chi_3$ Borophene[35] show extremely high Li/Na storage capacities but have poor ion diffusion speed. Thus there is urgent need to explore excellent electrode materials which can offer fast ion diffusion and high storage capacity simultaneously.

As a well-known semiconductor, GaN has been widely used for variable applications.[42] Unlike traditional 2D materials, it is previously quite a challenge to prepare freestanding 2D GaN by routine exfoliation technologies, because its bulk phase naturally crystallizes in the non-layered wurtzite structure.[43] Nevertheless, large area 2D

GaN has successfully synthesized quite recently, utilizing the graphene encapsulation method.[44] Freestanding 2D GaN is experimentally characterized to stabilize in the buckled structure, with surface fully passivated by hydrogen atoms. Its buckled structure, as well as the stability of the 2D GaN, have been further verified by following computations.[45-47] Especially, it is predicted that, when the 2D GaN is in the monolayer state, it would show interesting physical properties.[48] Beyond these exotic properties, 2D GaN is also potential to be a superior battery electrode material, considering the fact that 2D GaN is environmentally friendly and possesses a very low mass density. However, 2D GaN has not been investigated on this aspect yet.

In this work, by using the GaN monolayer as the prototype system, we perform first-principles computations to investigate the feasibility of 2D GaN as the battery electrode material. The detailed computational methods are provided in the Supporting Information. We fully study Li/Na adsorption and diffusion processes. We find that the GaN monolayer shows quite excellent performances as the battery anode material. Especially, the diffusion barriers for Li and Na ions are greatly lower than known 2D electrode materials. This manifests extremely fast ion diffusion on the GaN monolayer. Moreover, the semiconducting GaN become metallic after adsorptions with Li and Na atoms, with quite high ion storage capacities. Our results indicate that, the GaN monolayer can simultaneously possess good storage ability and fast ion diffusion as a battery electrode material.

## 2 Structure of 2D GaN

Bulk GaN crystallizes in the tetrahedral-coordinated wurtzite structure, so its monolayer would naturally possess unsaturated dangling bonds when directly cleaved from the bulk. Former experiments and computations[44,48] have verified that, the GaN monolayer is the most thermodynamically stable when its surfaces are fully passivated by hydrogen. We have also performed formation-energy computations to show the energy preference of the GaN monolayer. The formation-energy of H-passivated GaN monolayer yields to be 0.91 eV lower than the unpassivated one, which agrees well with previous results.[44,48]

In Fig. 1(a)-(c), we show the 4 × 4× 1 supercell of the optimized GaN monolayer. It manifests a buckled structure under hydrogen passivation. The optimized in-plane lattice parameter for the unit cell is 3.17 Å, which is only 0.63% underestimated comparing its bulk phase (3.19 Å).[49] The relaxed bonding lengths for Ga-N, Ga-H, and N-H are found to be 2.01 Å, 1.80 Å, and 1.50 Å, respectively.

## 3 Adatom adsorption on GaN

To begin with, we indentify the most favorable adatom adsorption site on the GaN monolayer. As shown in Fig. 1(a) and (c), for the top and the bottom surfaces of GaN, we have considered four high symmetry adsorption sites (denoted as T1-T4, and B1-B4, respectively). The adsorption energies for these sites are calculated by using the formula:

$$E_{\mathrm{Ad}} = E_{\mathrm{Li/Na+GaNH_2}} - E_{\mathrm{GaNH_2}} - E_{\mathrm{Li/Na}} \qquad (1)$$

In formula (1), $E_{\mathrm{Ad}}$ is the adsorption energy. $E_{\mathrm{Li/Na+GaNH_2}}$ ($E_{\mathrm{GaNH_2}}$) is the energy of the GaN supercell after (before) adatom adsorptions. $E_{\mathrm{Li/Na}}$ denotes the energy for per Li/Na atom.

Figure 2(a) and (b) show the calculated Li/Na adsorption energies. One can observe that their adsorption energy curves show quite similar features. First, GaN shows positive adsorption energies for all the bottom surface sites while has negative adsorption energies for all the top surface sites. This shows that the adsorptions prefer to occur on the top surface of GaN. Second, among the top adsorption sites (T1-T4), T2 and T3 sites have the same adsorption energy (because adatoms at T3 site would automatically shift to the T2 site), which is lower than those at T1 and T4 sites. Therefore, the adsorption at the T2 site is the most energetically favorable. From our computations, the adsorption energy is -0.46 eV (-0.25 eV) for per Li (Na) atom. These values are comparable with some typical 2D electrode materials. For instance, in terms of adsorption energies, we have for $Sr_2N$ (-0.13 eV for Na[27]),$V_2C$ (-0.16 eV for Li[22]), $Mo_2C$ (about -0.43 eV for Li[24], about -0.64 eV for Na[24]), and $MoS_2$ (-0.60 eV for Li[30]). In this sense, GaN is promising for 2D electrode materials.

Here we study possible charge transfers during adsorptions. We find that Li and Na atoms transfer their most electrons to the GaN monolayer. The results are shown in Table 1. The amounts of transferred charges are larger than 0.8 *e* for both adatoms. Such charge transfers indicate that these adatoms are indeed chemically adsorbed and form chemical compounds with the GaN monolayer.

We continue to study the conductivity of GaN after Li/Na adsorptions. For this point, we compare the electronic structure of the GaN monolayer before and after adsorptions. In Fig. 3, we display the corresponding total density of states (DOS) and the optimized atomic configurations. One can find that, although the pristine GaN monolayer is a semiconductor, it would become metallic during the adsorptions. This is a quite desired condition as a battery electrode candidate. The sufficient charge transfer from adatoms is responsible for this semiconductor-to-metal transition. Similar phenomenon has also been observed in other semiconducting electrode materials such as silicene,[41] phosphorus[37,38,50,51] and GeS sheets[39,52].

# 4 Adatom diffusion on GaN

Next, we estimate the rate performance of the GaN monolayer when using as LIBs and NIBs electrodes. For this point, we have considered three migration paths (P1, P2, P3) between neighbored T2 sites. The pathways and calculated diffusion profiles are shown in Fig. 4. One observes that the energy barrier for P2 is the lowest among the three paths, which is 79 meV for Li and 22 meV for Na. Further calculations show that the Li/Na diffusion barrier can be even lower with increasing the adatom coverage. We have calculated layer-dependent ion diffusion barrier on the GaN monolayer. Based on our computations, the Li (Na) diffusion barrier drops to 43 meV (16 meV) for one-layer-coverage and 37 meV (11 meV) for two-layers-coverage. It is worth noting that, these diffusion barriers are lower than most 2D anode materials. This indicates extremely fast Li/Na diffusion on the GaN monolayer. We will make detailed discussions on this aspect in section 6.

## 5 Li/Na storage capacity on GaN

In this section, we focus on the Li/Na storage capacity on GaN. The adsorption energies are computed layer-by-layer. Here the computations are performed on the 4 × 4× 1 supercell of the GaN monolayer. The half-cell reactions can be described as:

$$GaNH_2 + xM^+ + xe^- \longleftrightarrow MGaNH_2 \qquad M/M^+ \text{ (M = Li, Na)} \quad (2)$$

Therefore, the average OCV ($V_{ave}$) can be calculated follows:

$$V_{ave} = (E_{GaNH_2} + xE_M - E_{MGaNH_2}) / xe \qquad (3)$$

where $E_{GaNH_2}$, $E_{MGaNH_2}$ and $E_M$ have been defined above. Here $x$ is the number of adatoms. The voltage profile, which is the plot of $V_{ave}$ *vs.* Li/Na concentration, can reflect the sequence of the layer-by-layer Li/Na adsorptions on the GaN monolayer, since each layered adsorption would show a plateau in the voltage profile. To obtain the voltage profile, we first calculate the layer-resolved adsorption energy ($E_{layer-ave}$), which follows:

$$E_{layer-ave} = (E_{M_{16n}Ga_{16}N_{16}H_{32}} - 16E_M - E_{M_{16(n-1)}Ga_{16}N_{16}H_{32}})/16 \qquad (4)$$

For the calculated $E_{layer-ave}$, the values of Li yield to be: -0.34 eV of the first layer, -0.19 eV of the second layer, -0.08 eV of the third layer, 0.11 eV of the fourth layer; and Na: -0.21 eV of the first layer, -0.07 eV of the second layer, 0.08 eV of the third layer.

Figure 5 shows the Li/Na voltage profiles. For Li, the calculated voltage is about 0.261-0.060 V. The three voltage plateaus corresponds to the 1st, the 2nd and the 3rd layer adsorptions of Li atoms. For Na, the voltage is about 0.157-0.050 V. It shows two effective plateaus which corresponds to the 1st and the 2nd layer of Na adsorptions. To be noted, the third plateau in Fig. 5(b) shows a negative value of voltage. This indicates the third layer of Na is unlikely to be adsorbed. For LIBs and NIBs, the average OCV for the GaN monolayer is 0.05-0.26 V. These values are among those of typical anodes, for instance 0.12-0.20 V for graphite,[30,53] 0.12-0.68 V for borophene,[34,35] 0.09-0.23 V for $Ca_2N$,[27] 0.14 V for $Mo_2C$,[24,25] and 0.06 V for $Nb_2C$.[26] Thus GaN is suitable to be applied as a LIB and NIB anode material.

According to the definition of equation (3), one can find that, the maximum capacity on the GaN monolayer corresponds to the Li/Na concentration beyond which the $V_{\mathrm{ave}}$ becomes negative. As shown in Fig. 5(a) and (b), the chemical stoichiometry for the maximum Li and Na adsorption is $Li_3GaNH_2$ and $Na_2GaNH_2$, respectively. Then we can calculate the maximum Li/Na storage capacity ($C_{\mathrm{m}}$) by using the equation:

$$C_{\mathrm{m}} = x_{\mathrm{m}} F / M_{\mathrm{GaNH_2}} \qquad (5)$$

Similar computations are also used in previous works.[24-26,33] The calculated Li (Na) capacity is 938.1 (625.4) mA h $g^{-1}$.

Here we examine the thermodynamic stability of the $Li_3GaNH_2$ and $Na_2GaNH_2$ system. Ab initio molecular dynamics (AIMD) simulations are adopted by using $4 \times 4 \times 1$ supercells of $Li_3GaNH_2$ and $Na_2GaNH_2$ at 323 K. Each time step is chosen as 2 fs. Figure 6 shows the structure snapshots of initial and final states for $Li_3GaNH_2$ and $Na_2GaNH_2$. After 1500 steps, both $Li_3GaNH_2$ and $Na_2GaNH_2$ can retain their structures, indicating they can have good stability near the room temperature.

We also note that, several typical semiconductors BN, AlN and InN have the same structure with GaN. Their 2D counterparts are also promising to be synthesized by similar graphene encapsulation method.[44] Based on equation (5), we can make an approximate estimation of the possible Li/Na capacity of BN, AlN and InN. If they can have the same Li/Na adsorption abilities with the GaN monolayer, the Li capacity can be 3166.1 mA h $g^{-1}$ for BN, 1915.1 mA h $g^{-1}$ for AlN, and 609.6 mA h $g^{-1}$ for InN. As to Na, the capacities are 2110.7 mA h $g^{-1}$ for BN, 1276.7 mA h $g^{-1}$ for AlN, and 406.4 mA h $g^{-1}$ for InN, respectively. So this category of 2D materials is quite promising to serve as high capacity electrode materials.

# 6 Comparison with typical 2D anode materials

The diffusion barrier (which determines the rate capability) and the storage ability are quite crucial to battery electrodes. To get a comprehensive judgment of the electrode performance of the GaN monolayer, the parameters of some typical 2D anode materials for LIBs and NIBs are provided for comparison in Table 2 and Table 3.

For LIBs (see Table 2), one can find that the GaN monolayer possesses the second largest specific capacity among the listed ones. Although the capacity for the GaN monolayer is 0.3-1.1 times lower than $\beta_{12}/\chi_3$ Borophene, the Li diffusion can be 7.2-7.5 times faster in the GaN monolayer. When compared with commercial anode graphite, the GaN monolayer is 1.5 times higher in Li capacity and 4.7-14.2 times faster in Li diffusivity. Moreover, both Li storage capacity and diffusion performance in the GaN monolayer are much better than other anodes including phosphorene, $VS_2$, and GeS. It also worthies noticing that, some anode materials including $Mo_2C$, $Nb_2C$, and $Ti_3C_2$ possesses similar or a bit lower diffusion barriers, but their storage capacities (320-542 mA h $g^{-1}$) are greatly lower than GaN (938 mA h $g^{-1}$). Therefore, GaN is quite suitable to be used as an LIBs anode with simultaneously realizing fast ion diffusion and excellent storage ability.

For NIBs, as shown in Table 3, both Na diffusion performance and storage ability in the GaN monolayer are much better than most typical 2D anode materials including $Ti_3C_2$, GeS, $MoS_2$, and $TiS_2$. We also note that, $Nb_2C$ and $Mo_2C$ show similar diffusion barrier with the GaN monolayer, but their Na storage capacities are 1.3-3.7 times lower. For other anode materials listed in Table 3, including silicene/graphene, $MoN_2$, and $\beta_{12}/\chi_3$ Borophene, one can observe that they show very high Na storage capacities but suffer quite poor Na diffusion ability (the diffusion barriers are as high as 110-560 meV); while the proposed GaN monolayer can be 5.0-25.5 times faster in Na diffusion. Remarkably, among available 2D anode materials with similar scale of Na diffusion barriers, the GaN monolayer almost possesses the highest Na storage capacity.

## 7 Summary

In summary, using the GaN monolayer as a prototype system, we have theoretically investigated the properties of the newly synthesized 2D GaN as potential battery electrodes. We find Li and Na atoms exhibit negative adsorption energies with sufficient charge transfers to the GaN monolayer, indicating their chemically stable adsorptions. The GaN monolayer becomes metallic after Li and Na adsorptions, which ensures good electronic conductivity as battery electrodes. The calculated average OCV is as low as 0.060 V for Li and 0.050 V for Na. This indicates GaN is suitable for the anode material. Remarkably, we find both the ion diffusion performance and storage capacity in the GaN monolayer are much higher than most known 2D anode materials. To be specific, the diffusion barrier is only 79 meV and the capacity is 938 mA h $g^{-1}$ for Li, and those for Na are found to be 22 meV and 625 mA h $g^{-1}$. Our results suggest that both fast ion diffusion and excellent storage ability can be expected in the proposed GaN electrode.

## Acknowledgments

This work is supported by the Special Foundation for Theoretical Physics Research Program of China (No. 11747152), Natural Science Foundation of Tianjin City (No.16JCYBJC17200), Research Project for High Level Talent of Hebei Province (No.A2017002020).

**Table 1** Bader charge analysis for the GaN monolayer before and after Li/Na adsorptions.

| | Average charge state | | | | | |
|---|---|---|---|---|---|---|
| | Ga | H(-Ga) | N | H(-N) | Li | Na |
| $Ga_{16}N_{16}H_{32}$ | 1.739 | 1.297 | 6.356 | 0.608 | - | - |
| $Ga_{16}N_{16}H_{32}$+Li | 1.741 | 1.313 | 6.336 | 0.621 | 0.137 | - |
| $Ga_{16}N_{16}H_{32}$+Na | 1.742 | 1.303 | 6.328 | 0.631 | - | 0.170 |

**Table 2** Diffusion barrier and theoretical maximum capacity ($C_M$) of typical 2D anode materials for LIBs. The data of commercial graphite is also provided for comparison.

| Materials | Diffusion barrier (meV) | $C_M$ (mA h g$^{-1}$) | References |
|---|---|---|---|
| 2D GaN | 79 | 938 | This work |
| $Mo_2C$ | 15-43 | 400-526 | [24,25] |
| $Nb_2C$ | 32 | 542 | [26] |
| $Ti_3C_2$ | 70 | 320 | [23,54] |
| Phosphorene | 130-760 | 433-649 | [37,38,50] |
| $VS_2$ | 220 | 466 | [30] |
| $MoS_2$ | 210 | 355 | [29,55] |
| GeS | 236 | 256 | [39] |
| $\beta_{12}/\chi_3$ Borophene | 600-660 | 1240-1984 | [35] |
| $MoN_2$ | 780 | 432 | [33] |
| Graphite | 450-1200 | 372 | [2,56,57] |

**Table 3** Diffusion barrier and theoretical maximum capacity ($C_M$) of typical 2D anode materials for NIBs.

| Materials | Diffusion barrier (meV) | $C_M$ (mA h g$^{-1}$) | References |
|---|---|---|---|
| 2D GaN | 22 | 625 | This work |
| $Nb_2C$ | 15 | 271 | [26] |
| $Mo_2C$ | 19-25 | 132 | [24,25] |
| $Ti_3C_2$ | 96 | 319 | [23] |
| GeS | 100 | 512 | [39] |
| Silicene/graphene | 110 | 730-954 | [41] |
| $MoS_2$ | 280 | 146 | [58] |
| $TiS_2$ | >300 | 339 | [59] |
| $\beta_{12}/\chi_3$ Borophene | 330-340 | 1240-1984 | [35] |
| $MoN_2$ | 560 | 864 | [33] |

## Figures and captions:

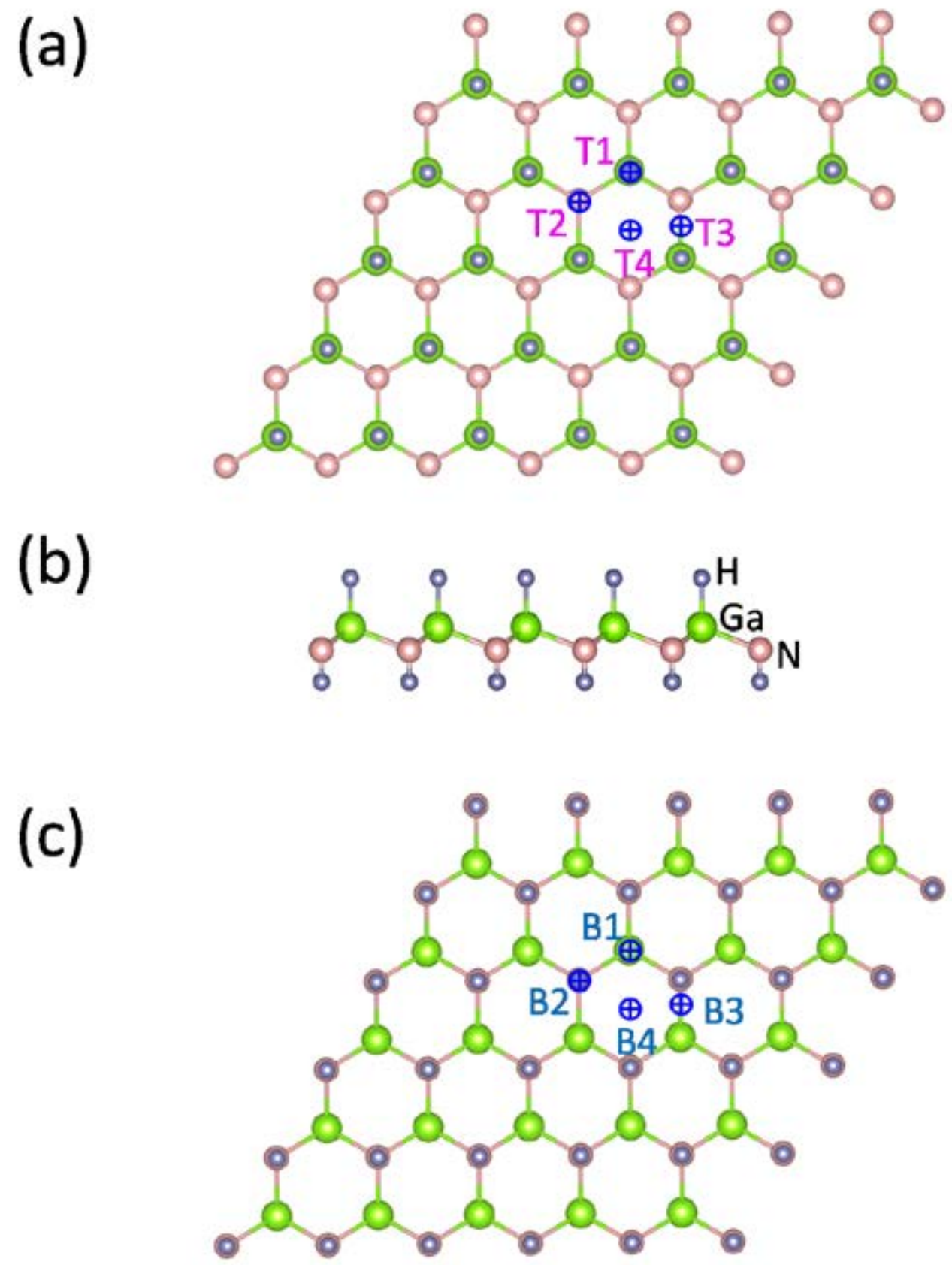

**Fig. 1** Atomic structures of the GaN monolayer in (a) top, (b) side and (c) bottom views. In (a) and (c), T1-T4 and B1-B4 represent the possible adsorption sites on the top and the bottom surfaces of the GaN monolayer.

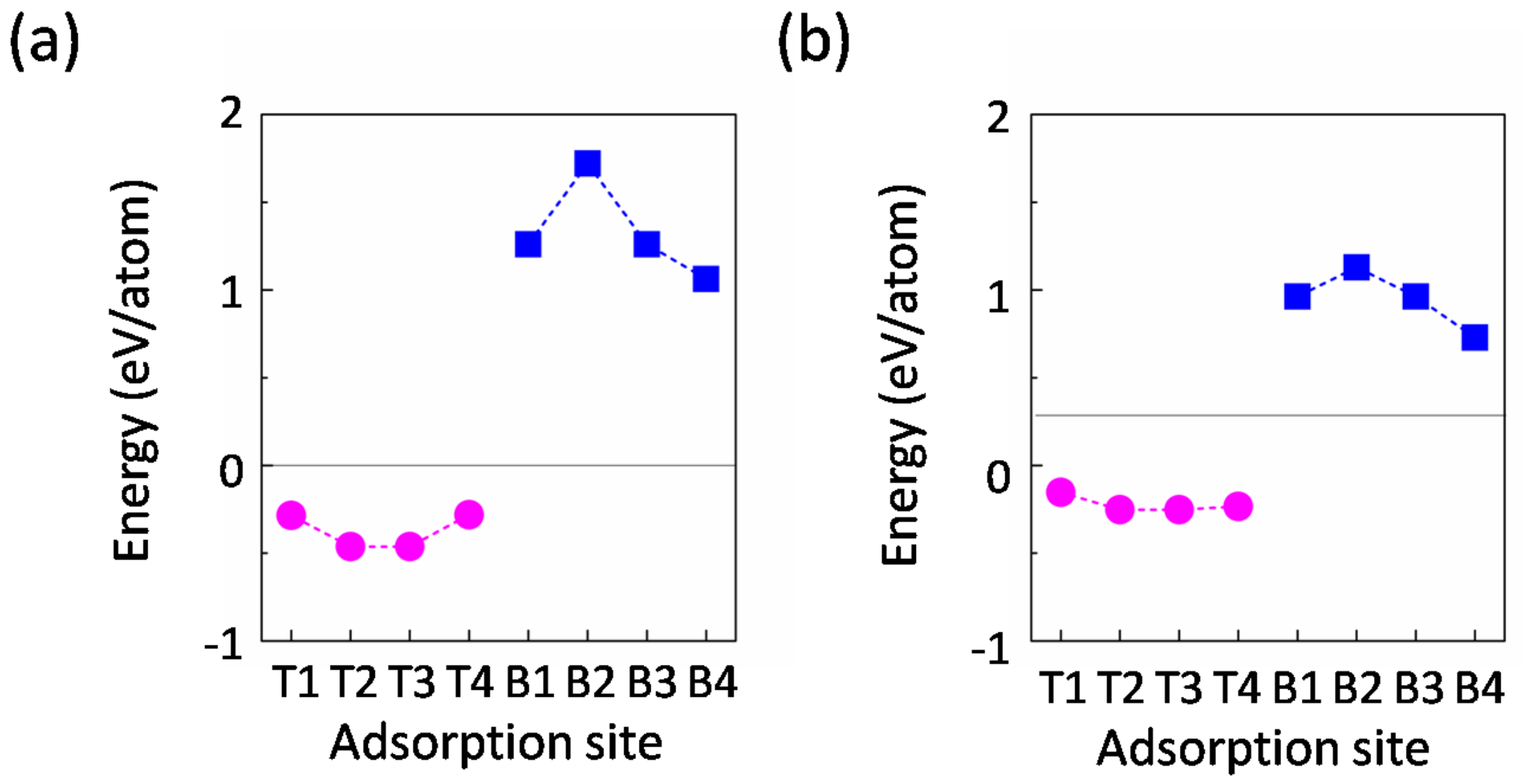

**Fig. 2** Adsorption energies for (a) Li and (b) Na atoms on the GaN monolayer at different adsorption sites.

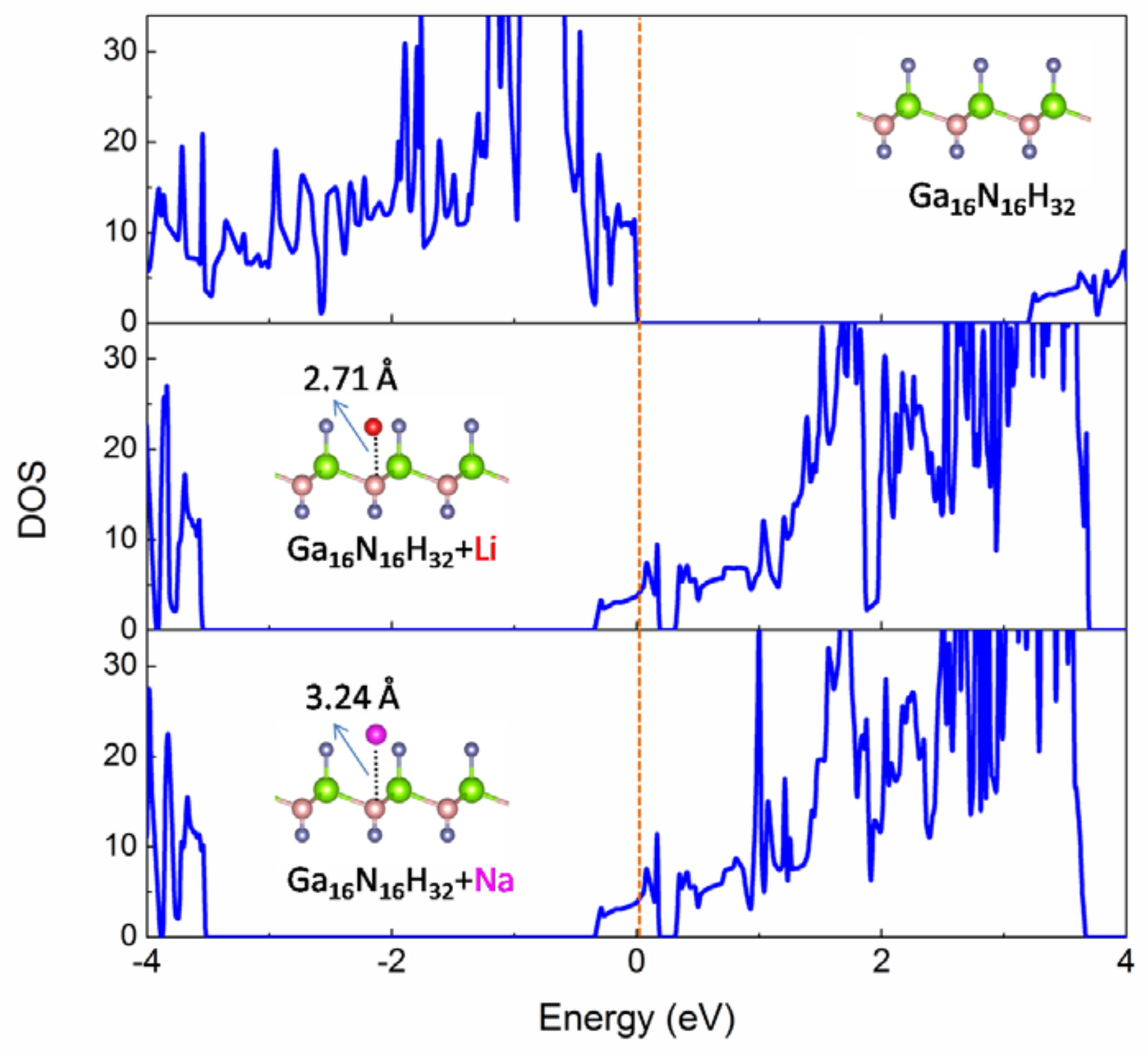

**Fig. 3** Total density of states (DOS) of the GaN monolayer before and after Li/Na adsorptions. Corresponding atomic structures are shown as the insets of the figure.

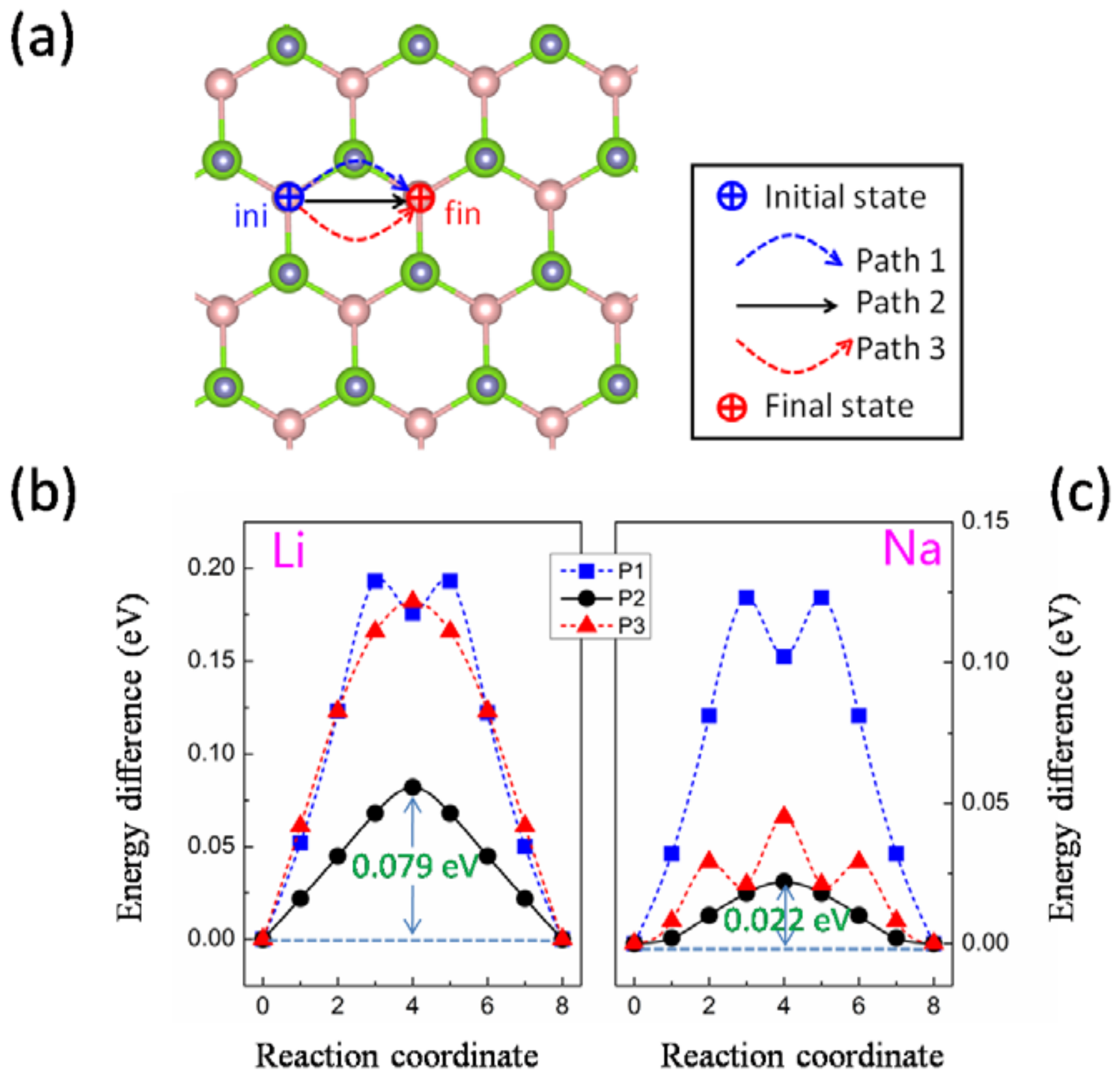

**Fig. 4** (a) Possible Li/Na ion migration paths on the GaN monolayer. Corresponding diffusion barrier profiles for Li and Na are shown in (b) and (c), respectively. In (b) and (c), the minimum Li/Na diffusion barriers are labelled.

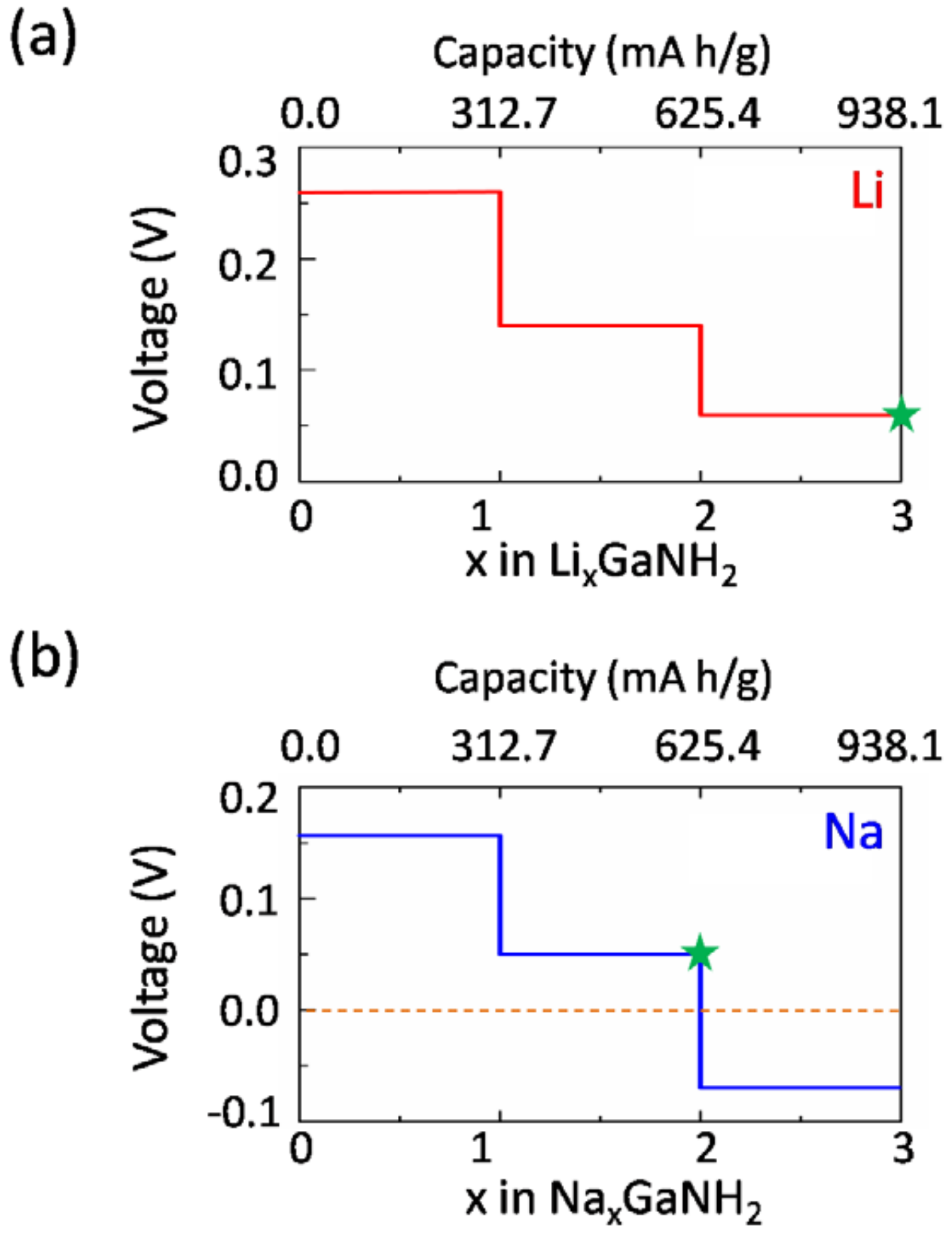

**Fig. 5** The calculated voltage profiles and storage capacities for (a) Li and (b) Na on the GaN monolayer. The maximum Li and Na storage capacities are indicated by the stars in (a) and (b).

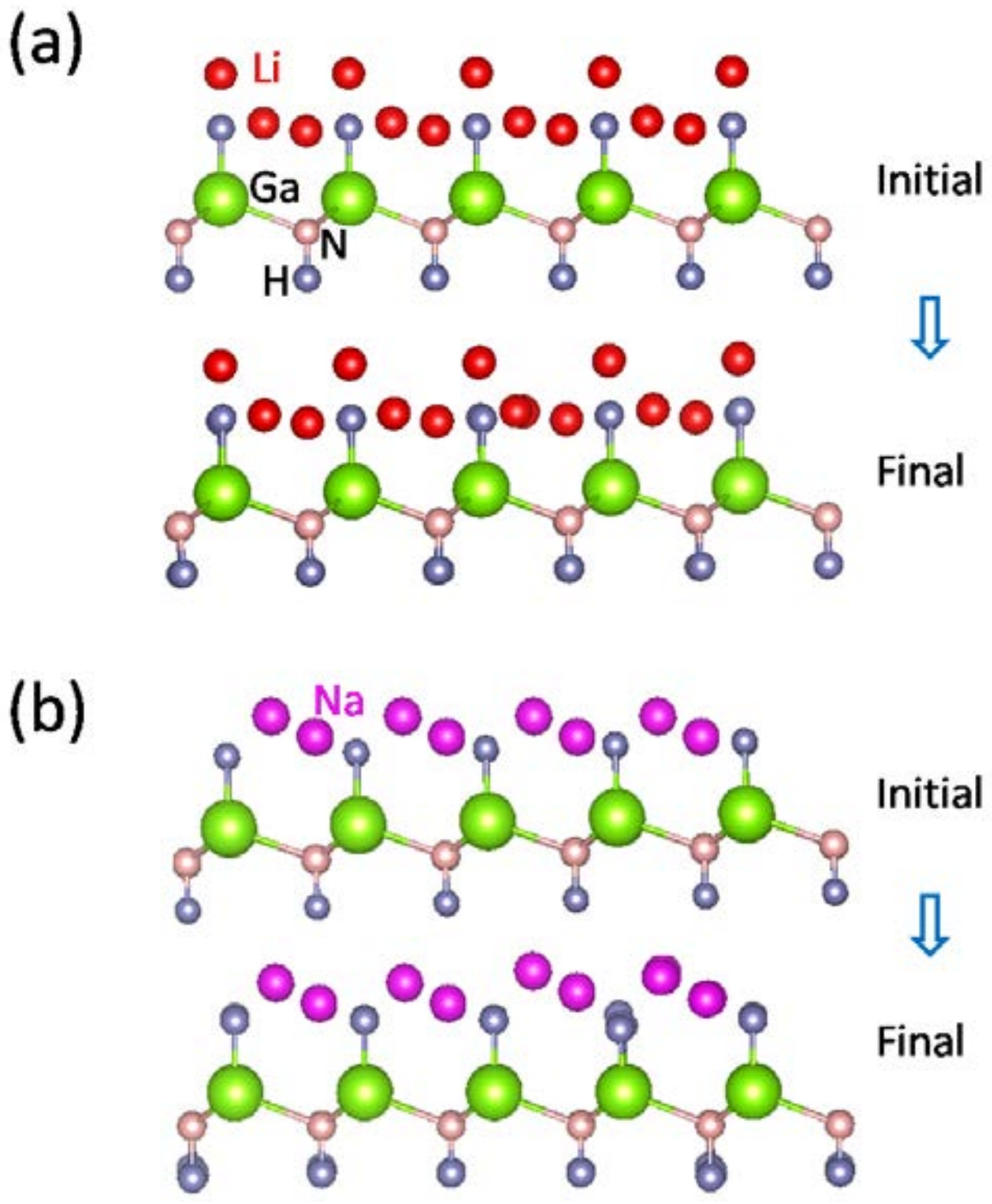

**Fig. 6** Comparison of structure snapshots between the initial and final states for (a) the $Li_3GaNH_2$ and (b) the $Na_2GaNH_2$ system after 3 ps of AIMD simulation at 323K.